\documentclass[twocolumn,showpacs,preprintnumber,amsmath,amssymb,nofootinbib]{revtex4}

\usepackage{epsfig}

\usepackage{graphicx}

\usepackage{bm}

\newcommand{\be}{\begin{equation}}
\newcommand{\ee}{\end{equation}}

\graphicspath{{./figure/}}

\begin{document}
\title{Extended Spherical Collapse and the Accelerating Universe}

\author{A. Del Popolo}
\email{adelpopolo@astro.iag.usp.br}
\affiliation{Astronomy Department, University of Catania, Italy}
\affiliation{Departamento de Astronomia, Universidade de S\~ao Paulo, Rua do Mat\~ao 1226,
05508-900, S\~ao Paulo, SP, Brazil}

\author{F. Pace}
\email{francesco.pace@port.ac.uk}
\affiliation{Institute of Cosmology and Gravitation, University of Portsmouth,
 Dennis Sciama Building, Portsmouth, PO1 3FX, U.K.}

\author{J. A. S. Lima}
\email{limajas@astro.iag.usp.br}
\affiliation{Departamento de Astronomia, Universidade de S\~ao Paulo, Rua do Mat\~ao 1226,
05508-900, S\~ao Paulo, SP, Brazil}


\date{\footnotesize{Received \today; accepted ?}}

\begin{abstract}

The influence of the shear stress and angular momentum on the nonlinear spherical 
collapse model is discussed in the framework of the Einstein-de Sitter (EdS) and $\Lambda$CDM models.
By assuming that the vacuum component is not clustering within the 
homogeneous nonspherical overdensities, we show how the local rotation and shear affects the linear density threshold 
for collapse of the non-relativistic component ($\delta_\mathrm{c}$) and its virial overdensity 
($\Delta_\mathrm{V}$). It is also found that the net effect of shear and rotation in galactic 
scale is responsible for higher values of the linear overdensity parameter as compared with the standard spherical collapse model (no shear and 
rotation). 

\end{abstract}

\maketitle

\section{Introduction}

Current analyses of high quality cosmological data are suggesting a 
 cosmic expansion history involving some sort of dark energy and a 
 flat spatial geometry in order to explain the recent accelerating 
 expansion of the universe \cite{SN,CMB,clust}.
 
 Among a number of possibilities to describe the dark energy (DE) 
 component, the simplest one is by means of a cosmological constant 
 $\Lambda$ (see \cite{reviews} for reviews), usually interpreted as 
 the vacuum energy density ($\rho_{\rm v}$) which acts on the Friedmann's 
 equations as a perfect fluid with negative pressure ($p_{\rm v} = -\rho_{\rm v}$).
 In the present cosmic concordance $\Lambda$CDM model, the overall 
 cosmic fluid contains non-relativistic matter (baryons + cold dark 
 matter, $\Omega_{\mathrm{nr}}=0.274$) plus a vacuum energy density 
 ($\Omega_{\mathrm{\Lambda}}=0.726$) that fits accurately the current 
 observational data and thus it provides an excellent scenario to 
 describe the present observed universe \cite{CM}.
 
 Nowadays, one of the most challenging problems in the so-called 
 $\Lambda$CDM  cosmology is to understand the role played by the 
 different cosmic components during the non-linear regime of 
 gravitational clustering and  how the many possible physical effects 
 contribute to determine the total mass of virialized halos (galaxy 
 and galaxy clusters). A popular analytical approach to study the 
 non-linear evolution of perturbations of dark matter (in the 
 presence of  a non-clustered dark energy (DE) is the standard spherical 
 collapse model (SSCM) proposed in the seminal paper of Gunn and Gott 
 \cite{GG72} and extended in subsequent papers \cite{SA}. The model describes how a spherical symmetric overdensity 
 decouples from the Hubble flow, slows down, turns around and 
 collapse.

In the last decade, the SSCM has been applied to study
density perturbation evolution and structure formation in presence  
of DE. However, when solving the density contrast ($\delta$) in the SSCM, 
the local shear ($\sigma$) and rotation ($\omega$) parameters are usually 
not taken into account. While the first assumption is correct, since for a sphere 
the shear tensor vanishes, the rotation term, or angular momentum is 
not negligible. A simple approach preserving spherical symmetry is to assume that 
the particles are described by a random distribution of angular 
 momenta such that the mean angular momentum at any point in space is 
zero \cite{WZ92N01}. Nevertheless, in any  proper extension of the SSCM both 
effects need to be considered \cite{EN2000} since shear induces 
contraction while vorticity induces expansion as expected from a centrifugal effect.

In this letter, we study the net physical effect of shear and rotation in the framework of an extended 
spherical collapse model (ESCM). We restrict our
analysis to the Einstein-de Sitter (EdS) and the flat $\Lambda$CDM background cosmologies.
For the $\Lambda$CDM model we assume the following cosmological parameters: $\Omega_{\mathrm{m}}=0.274$,
$\Omega_{\Lambda}=0.726$ and $h=0.7$. In particular, we discuss how the linear density threshold for 
collapsing non-relativistic component ($\delta_\mathrm{c}$) and its 
virial overdensity ($\Delta_\mathrm{V}$) change.  We recall that the change of these two parameters has a strong effect
on the mass function and other 
fundamental cosmological quantities. As a general result, it is also found that the extra terms appearing in the ESCM 
is responsible for higher values of the linear 
overdensity parameter at galactic scales as compared to the case without shear and rotation.

\section{Derivation of $\delta_{\mathrm{c}}$ and $\Delta_{\mathrm{V}}$}\label{sect:scm}

To begin with, let us now consider that the only clustering component in the cosmic medium is the cold dark matter. 
Following standard lines, the evolution of the overdensity $\delta$ is driven by a second order non-linear differential
 equation \cite{Pad,Pace2010}:

\begin{equation}\label{eqn:wnldeq}
 \begin{split}
  \delta^{\prime\prime}+\left(\frac{3}{a}+\frac{E^\prime}{E} \right)
\delta^\prime-\frac{4}{3}\frac{\delta^{\prime 2}}{1+\delta}-
  \frac{3}{2}\frac{\Omega_{\mathrm{m}}}{a^5E^2(a)}\delta(1+\delta)-&\\
  \frac{1}{a^2H^2(a)}(\sigma^2-\omega^2)(1+\delta)&=0\;,
 \end{split}
\end{equation}
where $a(t)$ is the scale factor and the scalars, $\sigma^2=\sigma_{ij}\sigma^{ij}$ and
$\omega^2=\omega_{ij}\omega^{ij}$, denote the shear and rotation terms, respectively ($\sigma_{ij}$ and $\omega_{ij}$,
are the shear and vorticity tensors). The quantity $\Omega_m$ is the present day value of the density parameter of the
DM component while the quantity $E(a)$  is defined by:
\begin{equation}\label{eq:e}
E(a)=\sqrt{\frac{\Omega_{\mathrm{m}}}{a^3}+\Omega_{\mathrm{\Lambda}}}\;,
\end{equation}
where $\Omega_{\Lambda}$ is the present day value of the vacuum density parameter (at $a=1$). 

Following Pace et al.~\cite{Pace2010} we now calculate the threshold for the collapse, $\delta_{\mathrm{c}}$, and the 
virial overdensity, $\Delta_{\mathrm{V}}$. We look for an initial density
contrast such that the non-linear equation diverges at the chosen collapse time. Once the initial overdensity is found,
we use this value as an initial condition in the linearised version of Eq.~\ref{eqn:wnldeq}.

The virial overdensity is readily obtained by using the definition
$\Delta_{\mathrm{V}}=\log(\delta_{\rm nl}+1)=\zeta(x/y)^3$, where $x=a/a_{\mathrm{ta}}$ is the normalized scale factor
and $y$ is the radius of the sphere normalized to its value at the turn-around.

In order to calculate the shear and vorticity terms in Eq.~\ref{eqn:wnldeq}, we first recall that
$\delta=\rho/\overline{\rho}-1=\left(a/R \right)^3-1$. 
As one may check, by inserting this expression into Eq.~\ref{eqn:wnldeq}, the evolution equation for the density 
fluctuations $\delta$ becomes \cite{Foss,EN2000,Ohta03}
\begin{equation}
\frac{d^2 R}{d t^2}= \frac{4}{3} \pi G \rho R - (\sigma^2-\omega^2)\frac{R}{3}= -\frac{GM}{R^2} - (\sigma^2-\omega^2)
\frac{R}{3},   
\end{equation}
which should be compared with the usual expression for the SCM with angular momentum (e.g.,~\cite{Pee93,Nuss01,Zuk10}):
\begin{equation}
\frac{d^2 R}{d t^2}= -\frac{GM}{R^2} +\frac{L^2}{M^2 R^3}= -\frac{GM}{R^2} + \frac{4}{25} \Omega^2 R,
\label{eqn:spher}
\end{equation}
where in the last expression we have replaced the momentum of inertia of a sphere, $I=2/5 M R^2$.
The previous argument shows that vorticity, $\omega$, is strictly connected to angular velocity, $\Omega$.

In the simple case of a uniform rotation with angular velocity ${\bm\Omega}=\Omega_z{\bf e}_z$, we have that
${\bm\Omega}=\omega/2$ (see also Chernin~\cite{Chern93}, for a more complex and complete treatment of the
interrelation of vorticity and angular momentum in galaxies).

It is also convenient to define the dimensionless $\alpha$-number as the ratio between the rotational and the
gravitational term:
\begin{equation}
 \alpha=\frac{L^2}{M^3 R G}\;.
\end{equation}
The above quoted ratio, $\alpha$, is of the order of 0.4, for a spiral galaxy like the Milky Way ($L \simeq 2.5 \times 
10^{74} g~cm^2/s$; $R \simeq 15$ kpc 
\cite{Ryd87,Cat96}), larger for smaller size perturbations (dwarf galaxies size perturbations) and smaller for larger 
size perturbations (for galaxy clusters the ratio is of the order of $10^{-6}$). 

Now, in order to integrate Eq.~\ref{eqn:wnldeq} one should determine how the extra term involving the difference
$\sigma^{2} - \omega^{2}$ depends on the density contrast. Based on the above outlined argument for rotation one may
calculate the same ratio between the gravitational and the extra term appearing 
in Eq.~\ref{eqn:wnldeq} thereby obtaining  
\begin{equation}
(\sigma^2-\omega^2)H_0^{-2}= -\frac{3}{2} \frac{\alpha\Omega_{\mathrm{m},0}}{a^3}\delta.  
\end{equation}

In the absence of a first principle workable expression, in what follows we will assume a more general power-law 
expression:

\begin{equation}
(\sigma^2-\omega^2)H_0^{-2}\equiv - b \delta^n/a^3,
\end{equation}
with the proviso that the constant parameters $b$ and $n$ can depend on the scale of the perturbations. In this concern,
 we remark that at the non-linear level, the gravitational term is $(1-\alpha)$ times smaller than for the case where 
rotation and shear is absent. This will have considerable effects on the linear and virial parameters of the
spherical collapse model, as we will see in the next section. In particular, the values of $b$ and $n$ can be 
calculated comparing the threshold of collapse, $\delta_{\rm c}$, in Sheth \& Tormen \cite{ST99} (later used to obtain
the Sheth \& Tormen \cite{ST01} mass function) with the $\delta_{\rm c}$ parameter which is obtained from
Eq.~\ref{eqn:wnldeq}. In this way, we have obtained $n=1$ and $b=0.157$, for galaxy size perturbations.

\begin{figure*}[!ht]
\centering
\includegraphics[angle=-90,width=0.39\hsize]{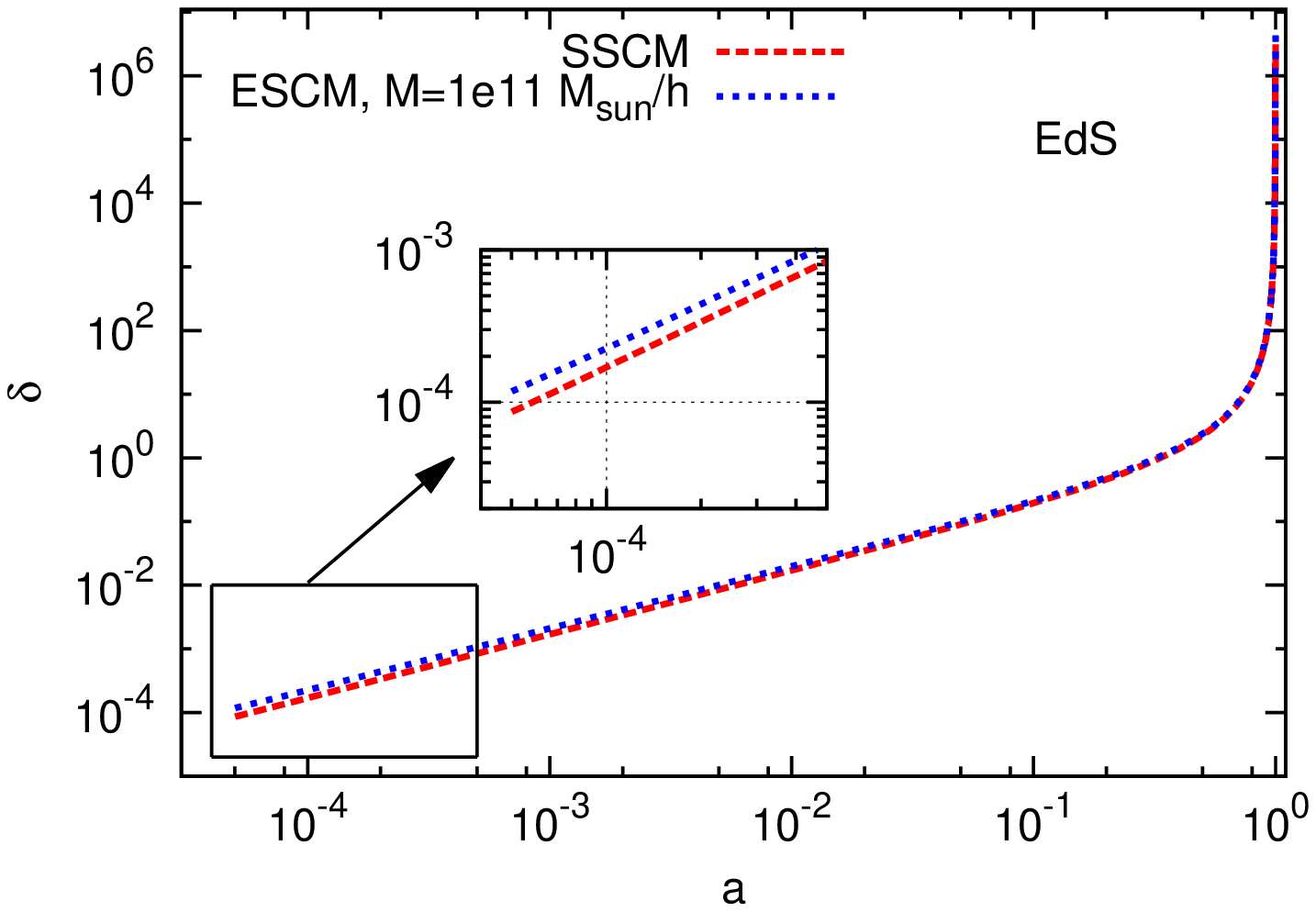}
\includegraphics[angle=-90,width=0.39\hsize]{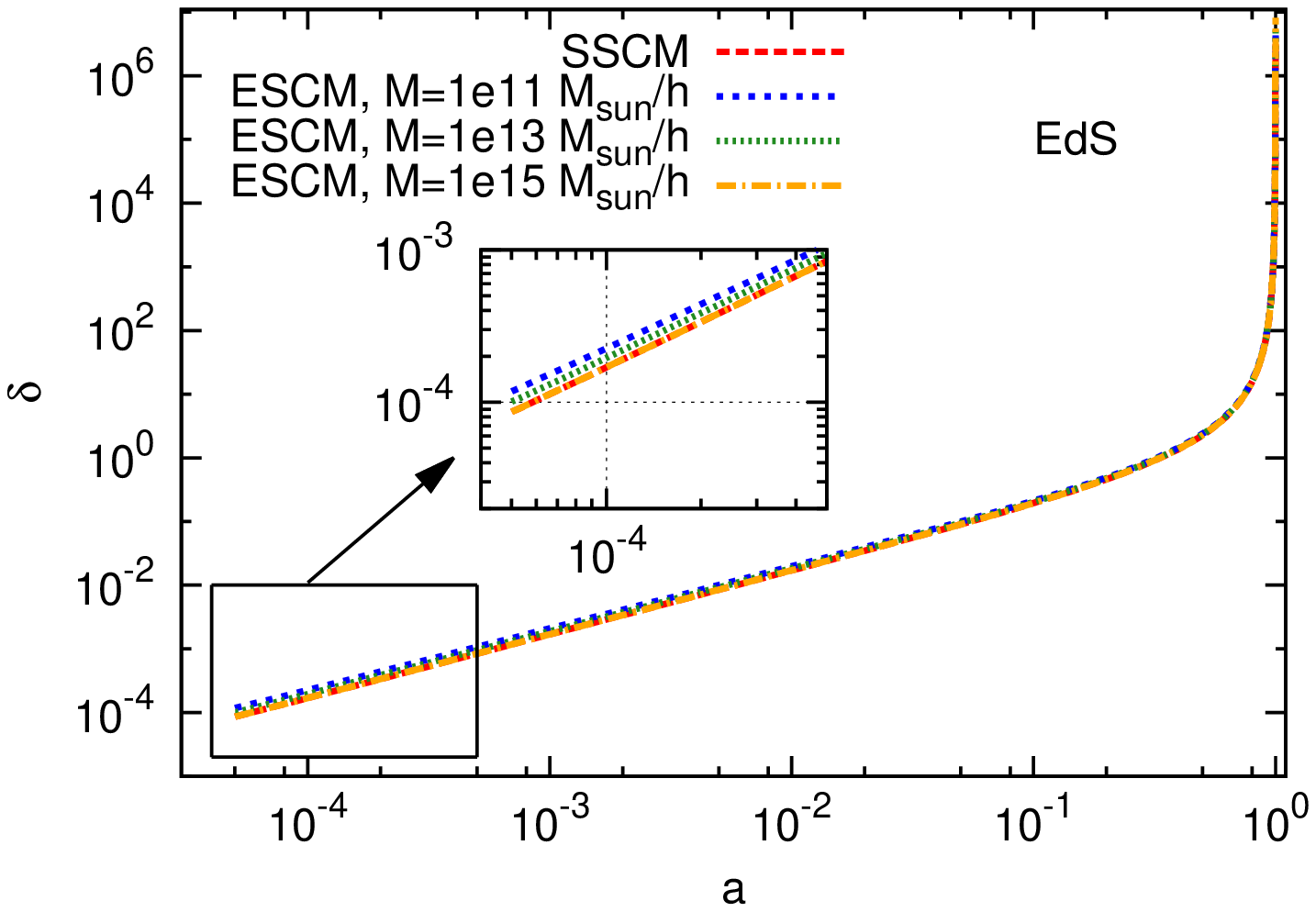}
\includegraphics[angle=-90,width=0.39\hsize]{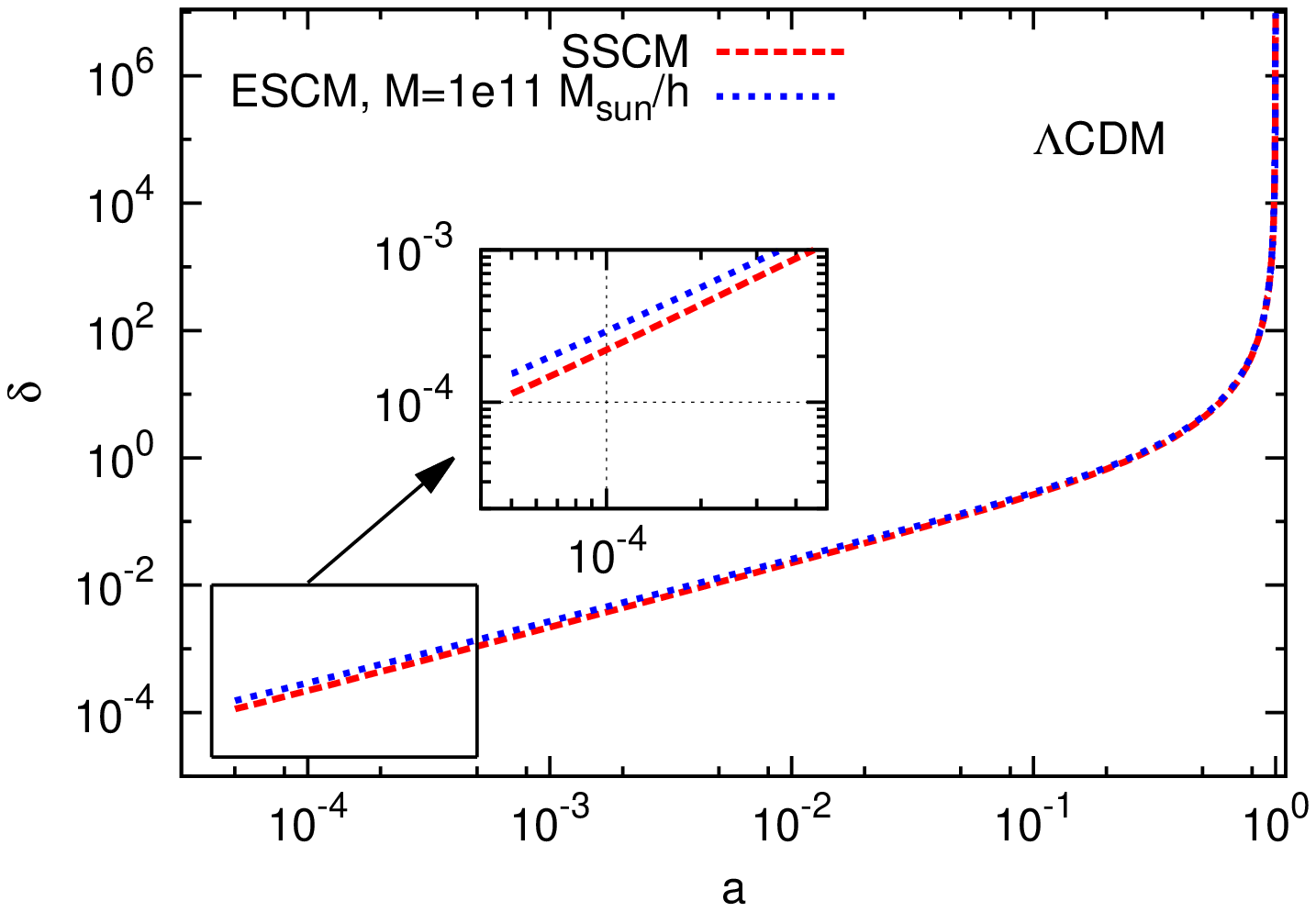}
\includegraphics[angle=-90,width=0.39\hsize]{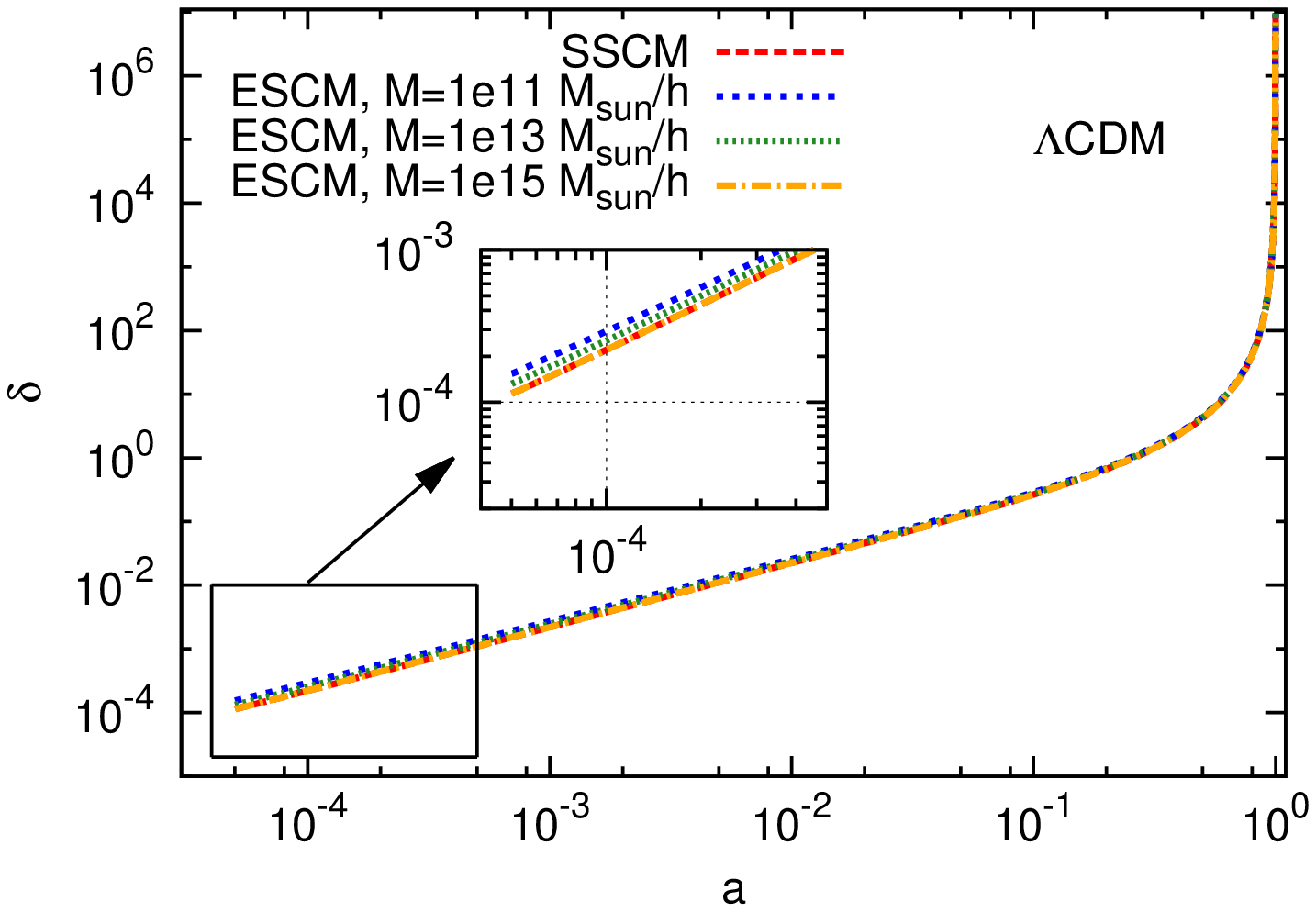}
\caption{The overdensity $\delta$ as a function of the scale factor $a$. On the left panels we compare the
evolution of the perturbation for the SSCM (red dashed curve) and ESCM (blue short-dashed) of
$10^{11}~M_{\odot}/h$ in the framework of an EdS and flat $\Lambda$CDM cosmologies. On the right panels we show similar 
plots but now for objects with different mass scales. The blue dashed curve shows results for a $10^{11}~M_{\odot}/h$
object (galaxy), the green short-dashed curve a $10^{13}~M_{\odot}/h$ (galaxy group)
while the orange dot-dashed curve represents a $10^{15}~M_{\odot}/h$ object (galaxy cluster). The images appearing the
above plots display a zoom on the initial overdensity necessary for the collapsing halo at redshift $z=0$. Note that the
initial overdensity of the ESCM halos need to be higher regardless of the background cosmology.}
\label{fig:deltaNL}
\end{figure*}

\section{Basic Results}\label{sect:Some basic results}
In this section we discuss some physical consequences of the extended spherical collapse model discussed here. In 
particular, we obtain the linear overdensity parameter $\delta_{\mathrm{c}}$ and the virial overdensity
$\Delta_{\mathrm{V}}$.

In Figure ~\ref{fig:deltaNL} (four plots) we show the evolution of the overdensity $\delta$ as a function of the scale 
factor assuming EdS and flat $\Lambda$CDM background cosmologies. In the left panels we compare the time evolution of
$\delta$ for the ESCM (short dashed blue curve) and the SSCM cases (red dashed line). The ESCM halo has a mass of
$10^{11}~M_{\odot}/h$. Since we want that both halos (spherical and non-spherical) should collapse at present time
($a=1$), the curves perfectly overlap in the non-linear regime. Therefore differences between the two considered models
must take place at very early times, reflecting therefore in the different initial conditions. For instance, whether
the collapse process is delayed in the case of ESCM halos (in comparison to the SCM description), one should expect
that the initial overdensity must be higher in order to have the same efficiency. This is indeed the case, as we can see
in the zoom panel on the left plot. In order to collapse at the same time of the non-rotating sphere, the initial
overdensity has to be higher. For the non-rotating sphere we have $\delta_{\mathrm{i}}\approx 8.6\times 10^{-5}$ while
for the rotating sphere we have $\delta_{\mathrm{i}}\approx 1.2\times 10^{-4}$, with an increase therefore of
approximately 28\%.

It should be noticed that in the right panels we show how the overdensities evolve for different masses of the 
corresponding halos (SSCM and ESCM) for the same background cosmologies. As expected, the influence of any departure 
from spherical symmetry decreases with the increase of the mass. For all practical purposes, we see that for scales
of the order of $10^{15}~M_{\odot}/h$ the solutions of the SCM are recovered. 

\begin{figure*}[!ht]
 \centering
 \includegraphics[angle=-90,width=0.39\hsize]{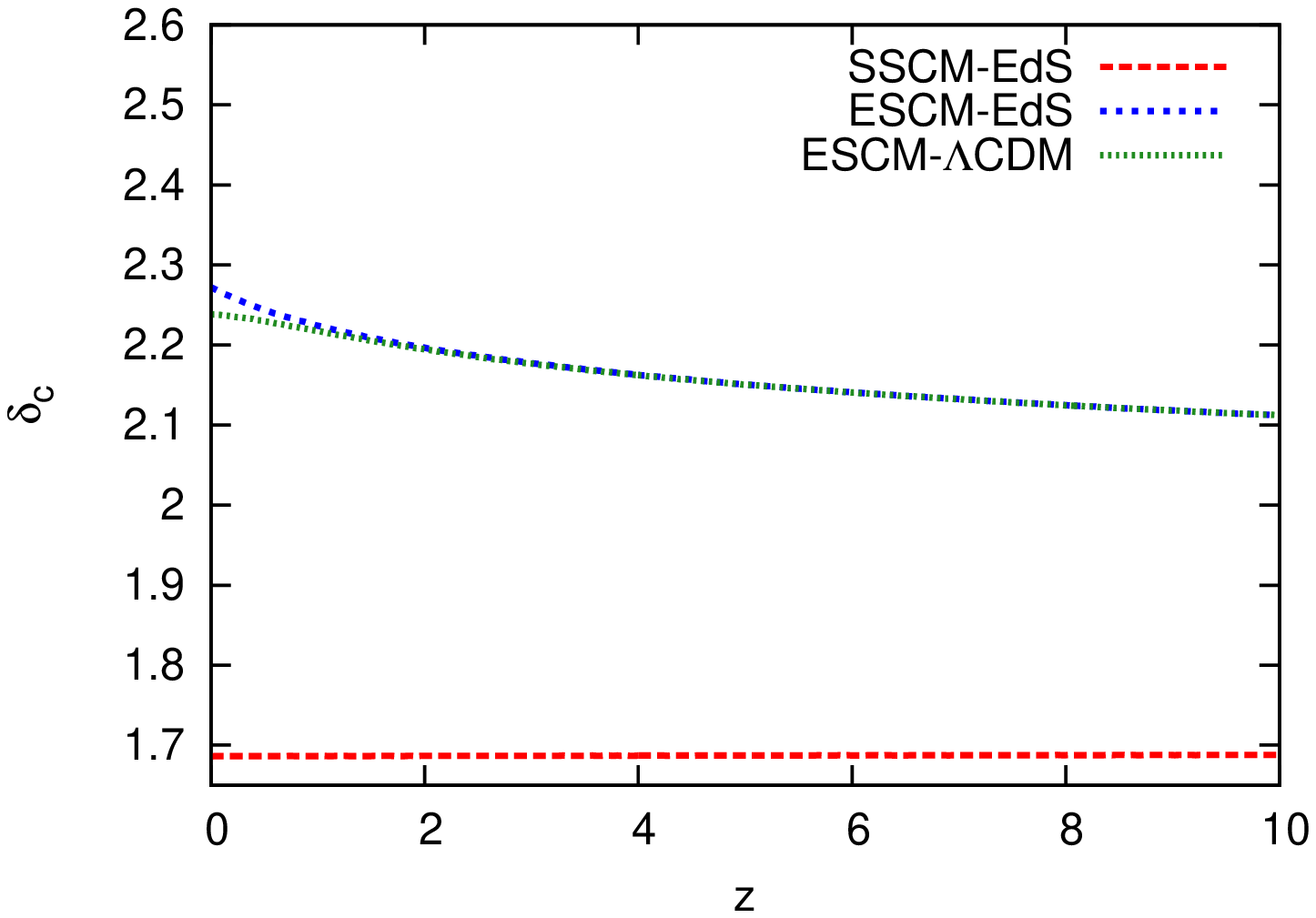}
 \includegraphics[angle=-90,width=0.39\hsize]{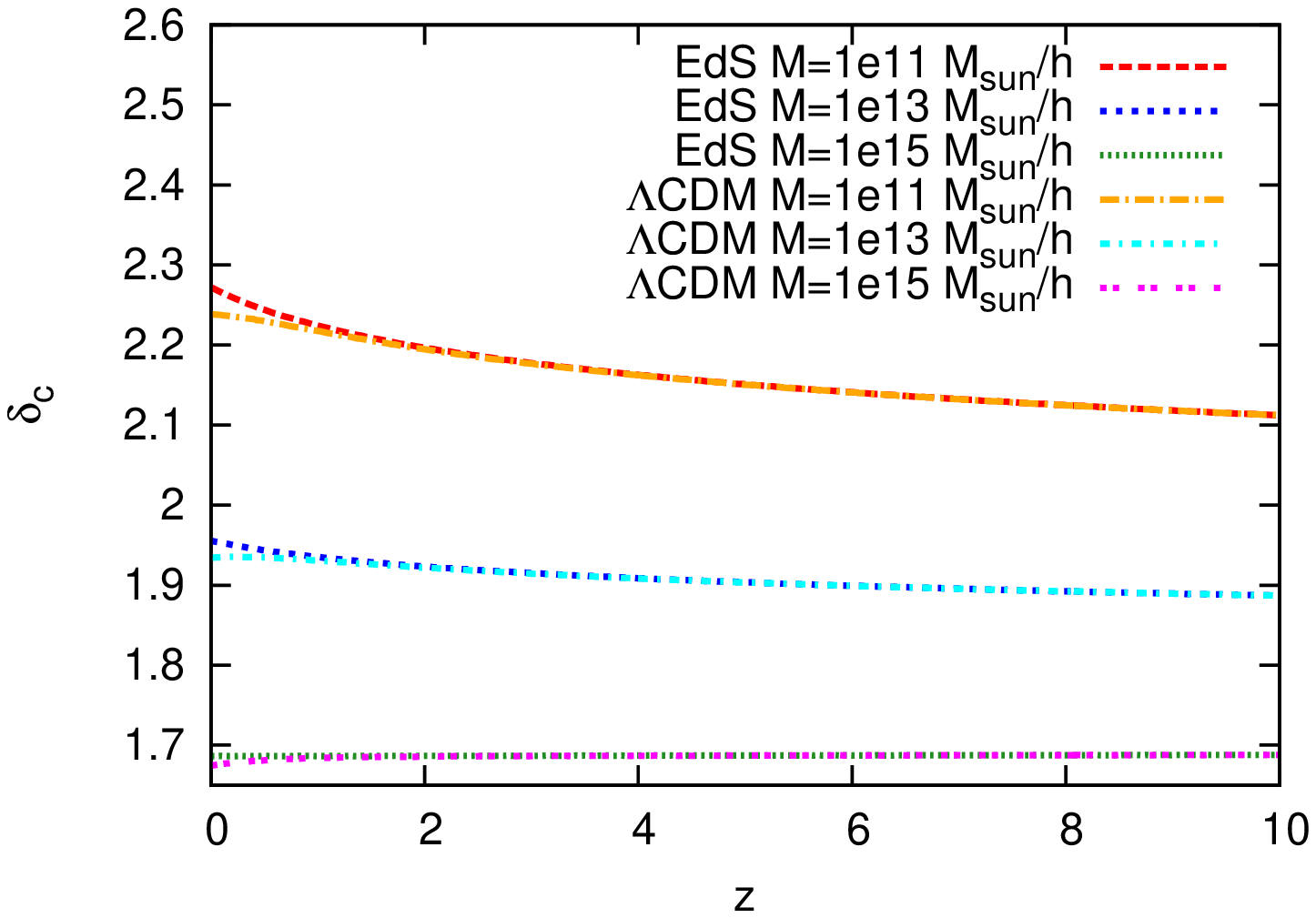}
 \includegraphics[angle=-90,width=0.39\hsize]{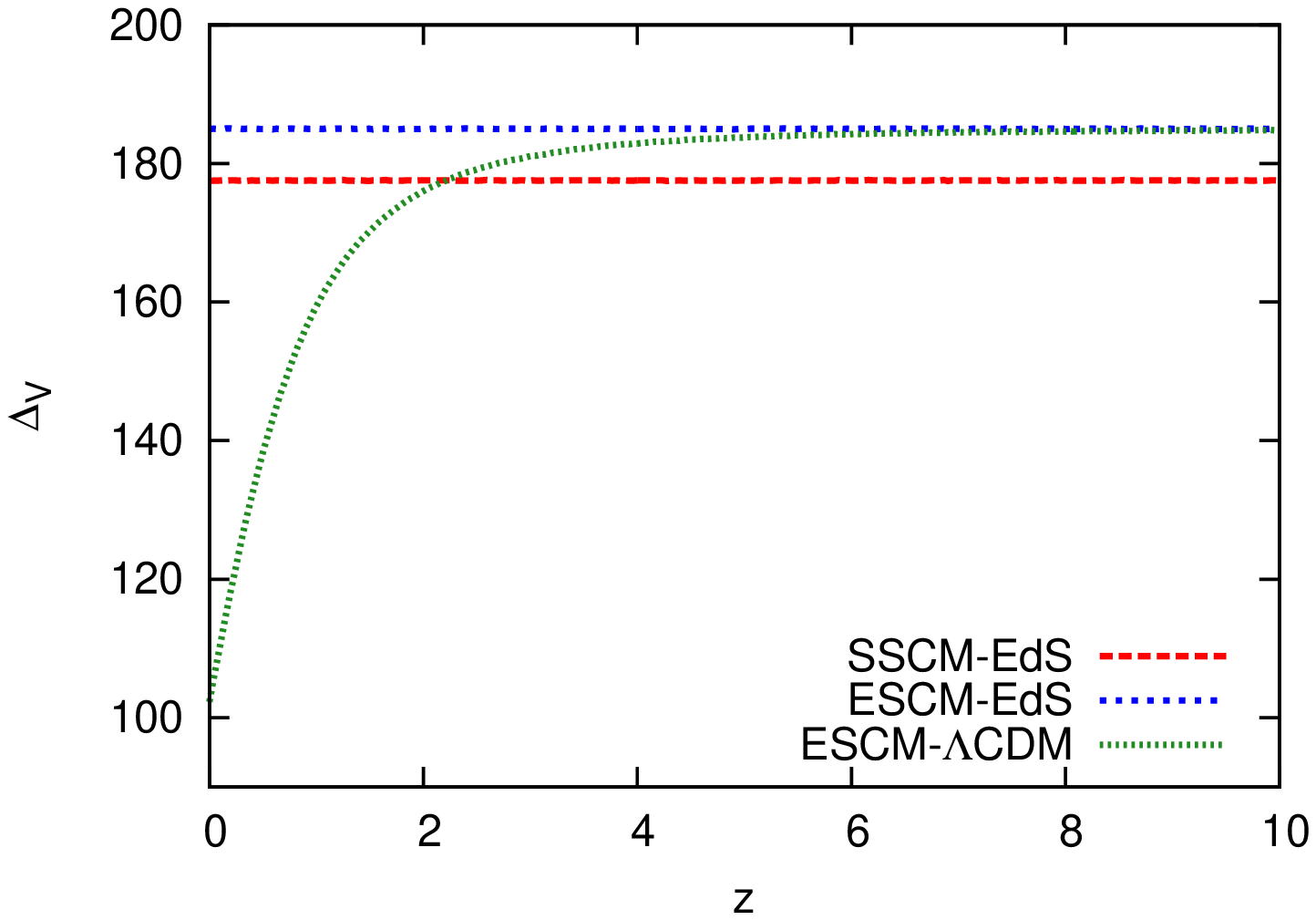}
 \includegraphics[angle=-90,width=0.39\hsize]{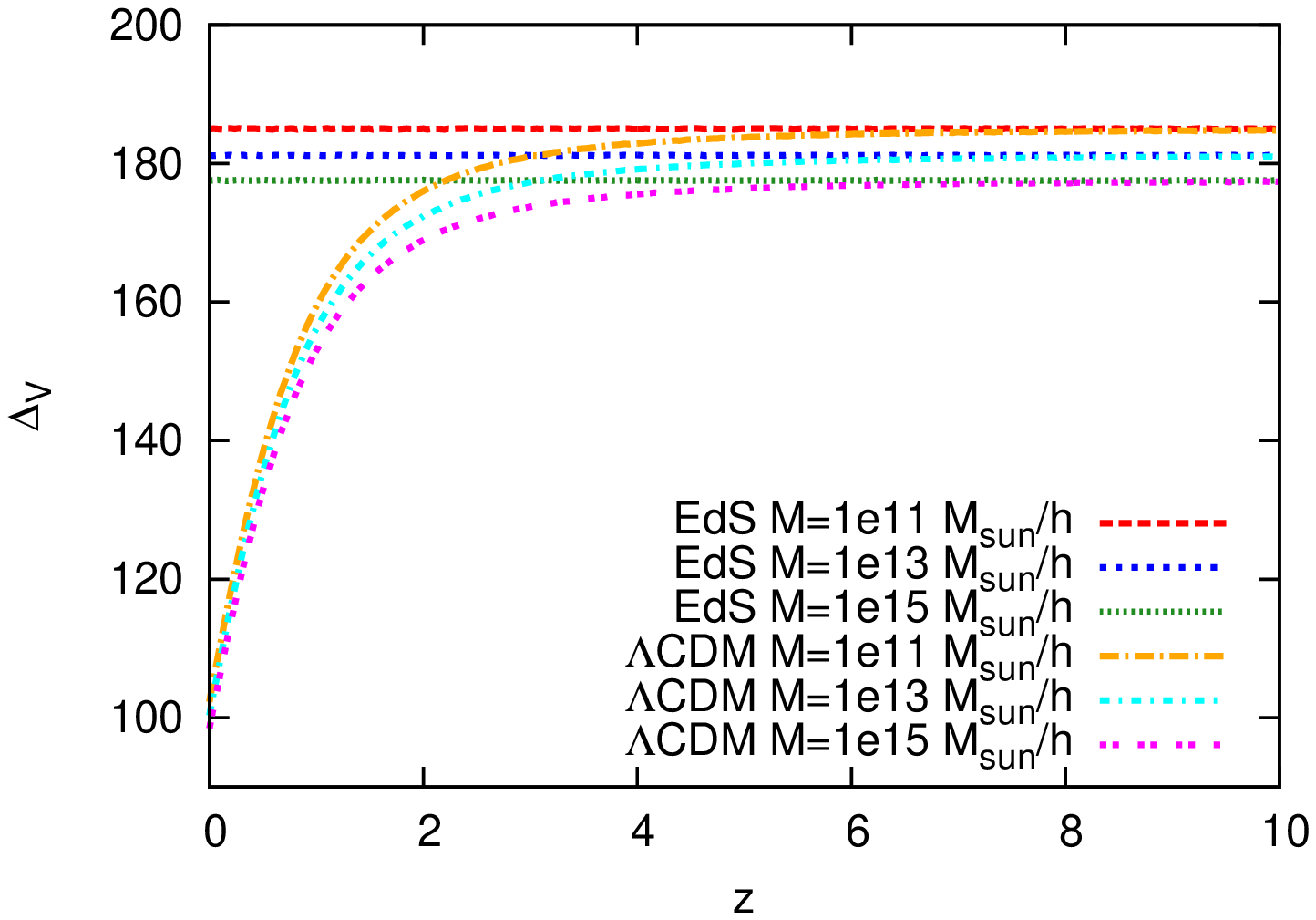}
 \caption{Upper (lower) panels: evolution of $\delta_{\mathrm{c}}$ ($\Delta_{\mathrm{V}}$) with respect to the redshift
$z$ for the EdS and the $\Lambda$CDM models. The left panels show the time evolution for both parameters at galactic
scale ($10^{11}~M_{\odot}/h$). The red curve represents the solution for the EdS model in the non-rotating case while
the blue and the green curves represent the EdS and the $\Lambda$CDM model when rotation is included. On the right
panels we compare the time evolution for three different masses, $10^{11}$, $10^{13}$ and $10^{15}~M_{\odot}/h$.
Different colours and line-styles correspond to different masses and different cosmological models: red dashed (orange
dot-dashed) curve represents a halo of $10^{11}~M_{\odot}/h$ in a EdS ($\Lambda$CDM) cosmological background, blue
short-dashed (dot-short-dashed cyan) curve represents a halo of $10^{13}~M_{\odot}/h$ for an EdS ($\Lambda$CDM) model,
while the green dotted (magenta) curve stands for an object of $10^{15}~M_{\odot}/h$ in an EdS ($\Lambda$CDM) model.}
 \label{fig:delta}
\end{figure*}

In Fig.~\ref{fig:delta} (4 plots), we show the evolution of the linear overdensity parameter $\delta_{\mathrm{c}}$
(upper panels) and of the virial overdensity $\Delta_{\mathrm{V}}$ (lower panels) for the same EdS and $\Lambda$CDM 
cosmologies. In the
left panels, the analyses based on the ESCM are restricted to a halo of $10^{11}~M_{\odot}/h$ since for galactic masses 
the effect will be
enhanced, while on the right panels we consider also the effect of distinct masses. As before, we concentrate our 
analyses to three different
mass scales: galactic  ($\approx 10^{11}~M_{\odot}/h$), groups ($\approx 10^{13}~M_{\odot}/h$) and clusters
($\approx 10^{15}~M_{\odot}/h$). As expected from the analysis of Fig.~\ref{fig:deltaNL}, with the growth of the mass
the effect of the extra term in the ESCM becomes negligible, and we recover the same values of the SSCM case. It is 
also worth to notice that the results for the $\Lambda$CDM model reduce to the ones of the EdS model for sufficiently
high redshifts, since the influence of the cosmological constant becomes rapidly negligible. We will therefore
concentrate only on the analysis of the left panels. For the different line colours and styles, we remind to the caption
of the figure.

As expected, the $\delta_{\mathrm{c}}$ for the ESCM is $\approx 40\%$ higher than for the  SSCM case and it decreases 
towards high redshifts, since the effect of the extra term becomes smaller. For the EdS model, $\delta_{\mathrm{c}}$
decreases from a value of $\approx 2.3$ at $z=0$ to $\approx 2.1$ at $z=10$. As expected, the linear overdensity
parameter for the $\Lambda$CDM model is smaller than the EdS one. This is understood by taking into account that if we
want to have the same number of structures now, we need to have a faster growth of structures to overcome the influence
of the cosmological constant. This translates into a lower $\delta_{\mathrm{c}}$.

In the lower panels we compare the behaviour of $\Delta_{\mathrm{V}}$ in the SSCM approach with the one predicted by
the ESCM description. The red dashed (blue short dashed) curve show the standard and the extended results for an EdS model while the green 
dotted curve represents a $\Lambda$CDM model. It is clear that the ESCM description affects also the virial overdensity parameter. In particular, we see that
$\Delta_{\mathrm{V}}$ is always constant in time for the EdS model. However, with the extra term its value increases
reaching $\Delta_{\mathrm{V}}\approx 185$, about 4\% higher than the standard result. The curve for the $\Lambda$CDM
model approximates the EdS at high redshifts, as expected. Once again higher masses are less affected by the ESCM
correction term (lower right panel).

\section{Conclusions}\label{sect:conc}

In this letter we have discussed how shear and rotation affect the standard spherical collapse model. The net effect 
of such quantities which is $\propto$ ($\sigma^{2} - \omega^{2}$) has been phenomenologically described by a power law 
on the density contrast depending on two parameters ($b$ and $n$). It was also shown that the values of $b$ and $n$ 
can be calculated by comparing the threshold of collapse, $\delta_c$, as discussed in Sheth \& Tormen \cite{ST99}, with 
the $\delta_{\rm c}$ value which is directly obtained from Eq.~\ref{eqn:wnldeq}. We have focused our discussion on
the influence of such an extra term on the spherical collapse parameters $\delta_{\mathrm{c}}$ and
$\Delta_{\mathrm{V}}$. As it should be expected, the extra term slows down the collapse, and, as such, higher values for
the initial perturbations are required in order to have a collapse at the same time of a spherical collapsing sphere.
It is also found that the extra term contribution is more important for galactic scales so that 
its contribution becomes negligible at high masses (galaxy clusters). This is shown explicitly in Figure ~\ref{fig:deltaNL}.

Finally, in Figure \ref{fig:delta} we have numerically evaluated and compared the evolutionary behaviour of both the ESCM and SSCM  
approaches. We have seen that both the linear and the non-linear virial overdensity in the extended spherical collapse 
model are enhanced with respect to the standard spherical case. Enhancements are more pronounced for $\delta_{\mathrm{c}}$
($\approx 40\%$), while for $\Delta_{\mathrm{V}}$ are only of the order of few percent.

These results reinforce the importance of a more complete and rigorous treatment involving the effects of shear and rotation 
at the late stages of the collapsing halo history mainly for the galactic scales. A more detailed article
including the calculations of the cumulative mass function will be published elsewhere.

\vspace{0.3cm} {\bf Acknowledgements:} ADP is partially supported by a visiting research fellowship from FAPESP (grant 
2011/20688-1), and wishes also to thank the Astronomy Department of S\~ao Paulo University for the facilities and 
hospitality. FP is supported by STFC grant ST/H002774/1, and JASL is also partially supported by CNPq and
FAPESP under grants 304792/2003-9 and 04/13668-0.

\end{document}